\documentclass[a4paper,11pt]{article}
\usepackage{aaskaiid}
\usepackage{orcidlink}

\title{Unveiling the roles of thermal and nonthermal processes in the ISM \& IGM structure formation and evolution of galaxies with SKAO }
\ShortTitle{Astrophysical processes in the ISM\& IGM  and evolution of galaxies}

\author[1,2,3]{Fatemeh S. Tabatabaei\orcidlink{0000-0002-0377-0970}}
\ShortName{Tabatabaei et al.} 
\author[4]{Masoumeh Ghasemi-Nodehi\orcidlink{0000-0001-6113-0317}}
\author[5]{Mark T. Sargent\orcidlink{0000-0003-1033-9684}}
\author[1]{Mahdiyar Mousavi-Sadr\orcidlink{0000-0002-5170-2534}}
\author[1]{Maryam Khademi\orcidlink{0000-0002-2865-0692}}
\author[3]{Eva Schinnerer\orcidlink{0000-0002-3933-7677}}
\author[6]{Eric J. Murphy\orcidlink{0000-0001-7089-7325}}
\author[7,8,9]{Elizabeth A. K. Adams\orcidlink{0000-0002-9798-5111}}
\author[10]{Isabella Prandoni\orcidlink{0000-0001-9680-7092}}
\author[11]{Cathy Horellou\orcidlink{0000-0002-3533-8584}}
\author[12]{Volker Heesen\orcidlink{0000-0002-2082-407X}}
\author[13]{Javier Moldon\orcidlink{0000-0002-8079-7608}}
\author[6,14]{Ilsang Yoon\orcidlink{0000-0001-9163-0064}}
\affiliation[1]{School of Astronomy, Institute for Research in Fundamental Sciences (IPM), PO Box 19395-5531, Tehran, Iran}
\emailAdd{ftaba@ipm.ir}
\affiliation[2]{Max-Planck Institut f\"ur Radioastronomie, Auf dem H\"ugel 69, 53121 Bonn, Germany}
\affiliation[3]{ Max-Planck-Institut f\"ur Astronomie, Department of Galaxies and Cosmology, K\"onigstuhl 17, D-69117 Heidelberg, Germany}
\affiliation[4]{State Key Laboratory of Radio Astronomy and Technology, Xinjiang Astronomical Observatory, CAS, 150 Science 1-Street, Urumqi 830011, China}
\affiliation[5]{ Institute of Physics, Laboratory of Astrophysics, \'Ecole Polytechnique F\'ed\'erale de Lausanne (EPFL), Observatoire de Sauverny, Versoix CH-1290, Switzerland}
\affiliation[6]{ National Radio Astronomy Observatory, Department of Astronomy, 520 Edgemont Road, Charlottesville, VA 22903, USA}
\affiliation[7]{ASTRON, Netherlands Institute for Radio Astronomy, Oude Hoogeveensedijk 4, 7991 PD Dwingeloo, The Netherlands}
\affiliation[8]{Kapteyn Astronomical Institute, University of Groningen, Postbus 800, 9700 AV Groningen, The Netherlands}
\affiliation[9]{SKA Observatory, SKA-Mid Science Operation Centre, 2 Fir St, Observatory, Cape Town 7925, South Africa}
\affiliation[10]{INAF - Istituto di Radioastronomia
Via Gobetti 101, 40129 Bologna (Italy)}
\affiliation[11]{Chalmers University of Technology, Department of Space, Earth and Environment, Onsala Space Observatory, SE-43992 Onsala, Sweden}
\affiliation[12]{University of Hamburg, Hamburger Sternwarte, Gojenbergsweg 112, 21029 Hamburg, Germany}
\affiliation[13]{Instituto de Astrof\'isica de Andaluc\'ia (IAA-CSIC), Glorieta de la Astronom\'ia s/n, 18008 Granada, Spain}
\affiliation[14]{University of Virginia, Department of Astronomy, Charlottesville, Virginia, United States}

\abstract{Investigating the thermal and nonthermal processes in the interstellar medium (ISM) and intergalactic medium (IGM)  is vital to understanding the evolution of galaxies over cosmic time. Resolved observations with SKA pathfinders show that the nonthermal processes, in which magnetic fields and cosmic rays are involved, can decelerate the formation of massive stars in strongly magnetized regions in nearby galaxies. They can also contribute to the onset of winds and outflows in galaxies. The effects of these processes are stronger at higher redshifts as a result of star formation activities.  The SKA Observatory will allow a major breakthrough by mapping the thermal and nonthermal processes in distant universe galaxies, shedding light on the role of the ISM and IGM in the evolution of galaxies. We demonstrate this by simulating the radio continuum and HI emission from local galaxies back to high redshifts.  Our simulations show that the AA4 surveys will make it possible to trace the thermal and nonthermal processes of the ISM in galaxies that are analogs to M51 and NGC6946, traced in continuum beyond cosmic noon (z=2-3) and the gas content traced by HI beyond z=1. Both simulations and precursor observations indicate the importance of nonthermal feedback at cosmic noon. }

\begin{document}
\maketitle
\newcommand{\actaa}{Acta Astron.} 
\newcommand{\araa}{ARA\&A} 
\newcommand{\aar}{A\&ARv} 
\newcommand{\aapr}{A\&ARv} 
\newcommand{\ab}{Astrobiol.} 
\newcommand{\aj}{AJ} 
\newcommand{\apj}{ApJ} 
\newcommand{\apjl}{ApJL} 
\newcommand{\apjs}{ApJSS} 
\newcommand{\ao}{Appl. Opt.} 
\newcommand{\apss}{Astro. \& Space Sci.} 
\newcommand{\aap}{A\&A} 
\newcommand{\aaps}{A\&AS.} 
\newcommand{\baas}{Bull. Am. Astron. Soc.} 
\newcommand{\caa}{Chinese A\&A} 
\newcommand{\cjaa}{Chinese J. A\&A} 
\newcommand{\cqg}{Class. Quantum Gravity} 
\newcommand{\gal}{Galaxies} 
\newcommand{\gca}{Geo. Cosmo. Acta} 
\newcommand{\icarus}{Icarus} 
\newcommand{\jcap}{JCAP} 
\newcommand{\jgr}{J. Geophys. Res.} 
\newcommand{\jgrp}{J. Geophys. Res. Planets} 
\newcommand{\jqsrt}{J. Quant. Spectrosc. Radiat. Transf.} 
\newcommand{\memsai}{Mem. SAIt} 
\newcommand{\mnras}{MNRAS} 
\newcommand{\nat}{Nature} 
\newcommand{\nastro}{Nat. Astron.} 
\newcommand{\ncomms}{Nat. Commun.} 
\newcommand{\nphys}{Nat. Phys.} 
\newcommand{\na}{New Astron.} 
\newcommand{\nar}{New Astron. Rev.} 
\newcommand{\physrep}{Phys. Rep.} 
\newcommand{\pra}{Phys. Rev. A} 
\newcommand{\prb}{Phys. Rev. B} 
\newcommand{\prc}{Phys. Rev. C} 
\newcommand{\prd}{Phys. Rev. D} 
\newcommand{\pre}{Phys. Rev. E} 
\newcommand{\prx}{Phys. Rev. X} 
\newcommand{\prl}{Phys. Rev. Let.} 
\newcommand{\psj}{Planet. Sci. J.} 
\newcommand{\planss}{Planet. Space Sci.} 
\newcommand{\pnas}{Proc. Natl Acad. Sci. USA} 
\newcommand{\procspie}{Proc. SPIE} 
\newcommand{\pasa}{PASA} 
\newcommand{\pasj}{PASJ} 
\newcommand{\pasp}{PASP} 
\newcommand{\rmxaa}{RMXAA} 
\newcommand{\sci}{Science} 
\newcommand{\sciadv}{Sci. Adv.} 
\newcommand{\solphys}{Sol. Phys.} 
\newcommand{\sovast}{Soviet Ast.} 
\newcommand{\ssr}{Space Sci. Rev.} 
\newcommand{\uni}{Universe} 

\section{Introduction}
In galaxy formation scenarios, feeding from the intergalactic medium (IGM) maintains cool gas reservoirs to fuel star formation and supermassive black hole accretion \citep[e.g.,][]{Kauf,Tacconi_2013,Walter_2020}. 
 A lack of cool gas, caused by different mechanisms such as excess feedback and starvation, is considered as the main cause of quenching of star formation in galaxies \citep[e.g.,][]{dave,peng}.  
However, there is evidence that ample amounts of cool gas exist in quenched
systems, but are, for some unknown reason, inefficient in forming
massive stars \citep{Gobat_18,gobat_20}. 
Hence, factors controlling gas against
collapse and accretion over cosmic time must be better understood.
Observations show that nonthermal pressures inserted by cosmic
rays and magnetic fields can be dynamically important \citep{beck07,Taba_08,Hassani,nasir}.
{Recent observations with the SKA precursors and pathfinders indicates enhanced energies of cosmic ray electrons (CREs) and magnetic fields at 1.5 < z < 3.5, supporting the importance of nonthermal processes at cosmic noon \citep{taba_25}.}
These processes, which can support cool gas
against accretion from the IGM \citep{Math,owen,gron} 
and collapse/fragmentation in the interstellar
medium \citep[ISM, e.g.,][]{pillai}
can decelerate the star formation rate \citep[SFR,][]{taba_18}. Therefore, it is insightful to dissect the thermal and nonthermal processes over cosmic time.
These processes are best studied  through mapping the radio
continuum (RC) emission in galaxies. 
Deep and resolved RC observations of distant galaxies (back to cosmic noon) are needed to
address the exact role of these astrophysical processes and the origin and efficiency of feedback. However, currently radio deep-field surveys have too low angular resolution and sensitivity to allow for the separation of the thermal and nonthermal emission in distant
galaxies. The advent of the Square Kilometre Array (SKA) and
its instrumental capabilities, combined with ground-breaking results
from ALMA, VLT/MUSE, JWST, and others, which trace different phases
of the gaseous ISM, will shed light on this subject.

In this chapter, we first provide a brief overview of the existing challenges and caveats in the standard galaxy-evolution framework (Sect. 2). We then present new perspectives on the role of astrophysical processes in the ISM and IGM, revealed through radio-continuum observations with SKA pathfinders and precursors over the past decade (Sect. 3). These results highlight the transformative potential of SKA observations for addressing the key questions outlined in Sect. 4, a conclusion further supported by our simulations (Sect.~5).
Throughout the chapter, we use H$_0$ = 67.4 km\,s$^{-1}$\,Mpc$^{-1}$, $\Omega_{m}$ = 0.31, and $\Omega_{\lambda}$= 0.68 (Planck Collaboration VI 2020).

\section{Feeding \& Feedback scenario and open questions}
Observations show that galaxies become redder and less luminous over cosmic time,
which is linked to the observed deceleration of the massive star formation rate \citep[SFR, e.g.,][]{Faber}. This is explained by a drop in the amount of cold gas that fuels star
formation through various mechanisms such as feedback and starvation \citep[e.g.,][]{Schaye,peng}. As such, the cold gas content of galaxies must decrease with time
at the same rate as that of star formation. 
%
%
Observations of the HI 21-cm line, which are mainly based on stacking analysis, show no evidence for a fast decline in atomic gas over cosmic time \citep[e.g.,][]{sini}. Analysis of stacked HI 21\,cm emission signal of the 11419 star-forming galaxies, which have an average stellar mass of M$_{\star}\simeq 10^{10}\, {\rm M}_{\odot}$, also shows only a factor of 3 higher HI mass at $z=1$ indicating a drop in the star-formation efficiency (SFE) over cosmic time \citep{Chowdhury}. More sensitive observations are required to shed light on the cosmic evolution of HI.  Nevertheless, the cosmic evolution in the SFE is also deduced through ALMA observations of molecular gas in the main-sequence star-forming galaxies \citep[e.g.,][]{scoville,tacconi_18,brown}. Observations in nearby galaxies show that there is often a considerable amount of quiescent cool gas available that is not forming stars \citep{Schinner} and that a drop in the SFE must have occurred in quenching galaxies \citep[e.g.,][]{Colombo}. 

Cosmological feedback models are mainly assuming thermal winds from supernova (SNe) and AGN \citep{Schaye,vecchia,oku} and radiative or mechanical AGN modes \citep[e.g.,][]{sij}. However, the efficiency of these kinds of SNe/AGN feedback in quenching star formation and galaxies are questioned by several studies \citep[e.g.,][]{ward} and are inconsistent with observations \citep[e.g.,][]{Suresh,maio2,kos,maio}. 
%
Theoretical studies show that cosmic rays can drive galactic winds in galaxies where star formation activity is low, so pure thermally-driven winds are not viable \citep{sike}.  
Cosmic rays also lead to cooler outflows with higher mass loss rates \citep{Giri,rath}. They may also play an important role in accelerating cold gas clouds that contain the bulk of the mass loss-rate \citep{Bruegen}. 
However, these theoretical studies must be confirmed by observation. For example, radio continuum observations show that the magnetic/cosmic-ray-driven winds effectively occur in star-forming regions in galaxies across a wide range of mass and star-formation properties \citep[][see Sect.~3]{Taba17,taba_22}.

Important questions are: What are the processes that prevent cold gas from forming stars? Why is the star formation efficiency reduced over cosmic time? What kind of feedback has been responsible for quenching galaxies?  A detailed study of astrophysical processes and energy/pressure balance in the ISM and the IGM is vital to
pinpoint factors controlling gas against collapse and accretion on different scales over cosmic time and to distinguish different origins of winds and outflows.

\section{Radio continuum observations: key to energy balance studies}
\begin{figure*} 
\centering
\includegraphics[width=0.7\hsize]{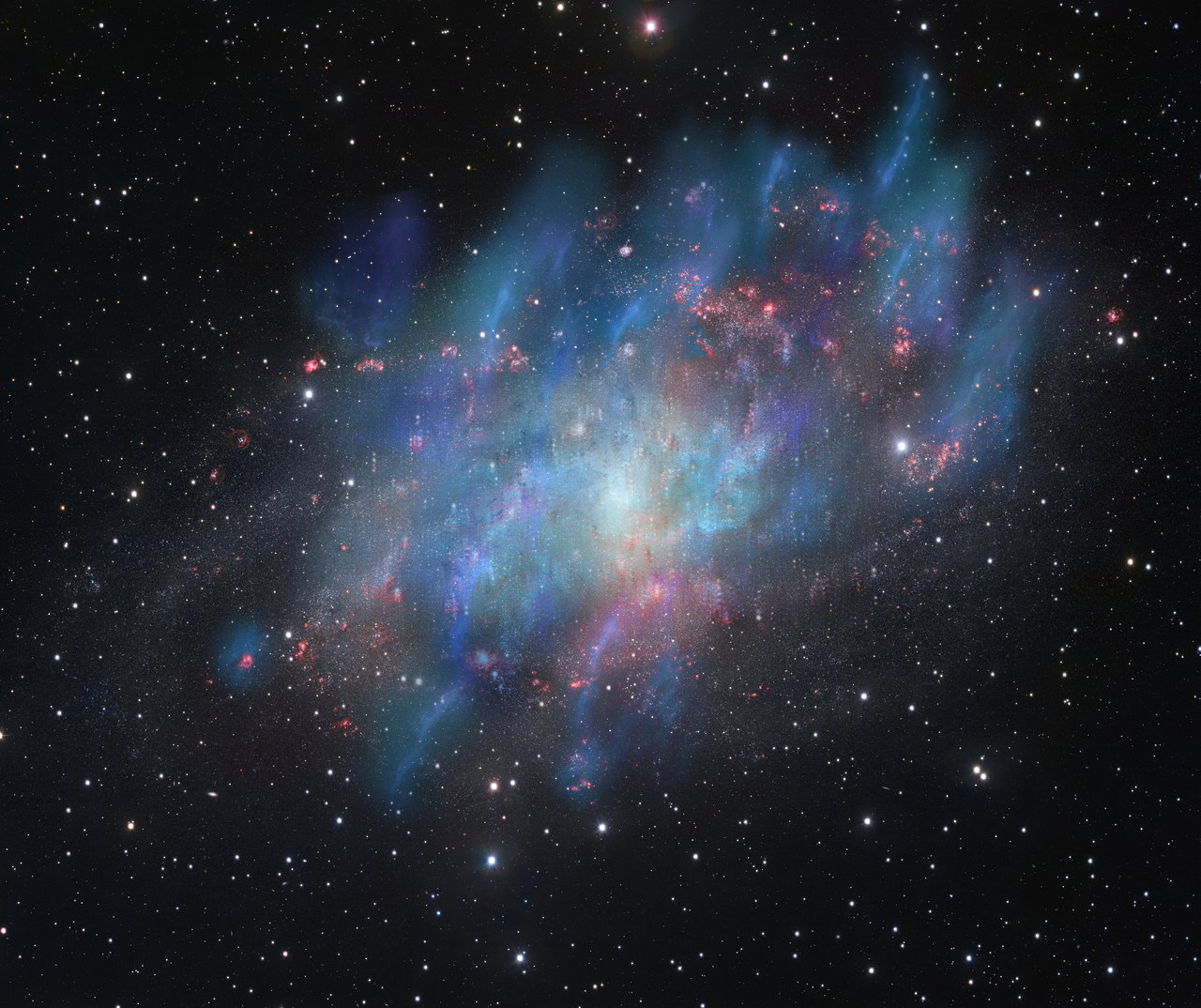}
      \caption{Artist's illustration of the {\it nonthermal feedback} \citep[due to cosmic rays in turbulent magnetic field,][]{taba_22} in star-forming regions (blue and green) superimposed on a VLT visible-light image (red and white). Credit: Institute for Research in Fundamental Sciences- IPM \& European Southern Observatory (ESO).}
\label{M33}
\end{figure*}

The RC emission from galaxies originates from different thermal and nonthermal physical processes in the ISM: The free-free continuum component emerges from the thermally ionized gas in HII regions and diffuse warm ionized medium. The synchrotron component is due to cosmic ray electrons (CREs)  injected in star-forming regions (supernova remnants) and propagated in the magnetized ISM. The power-law index of the nonthermal synchrotron spectrum, which is related
to the energy index of CREs, varies in
a galaxy \citep{taba_07,taba_13,Hassani,nasir} being flatter in star-forming regions and steeper in inter-arm
regions and outer disks. This indicates that CREs are more energetic closer to their birthplace in star-forming regions 
than in other places due to cooling mechanisms \citep[e.g.,][]{longair}.
Global surveys such as the radio follow-up survey of the Key Insights on Nearby Galaxies: A Far
Infrared Survey with Herschel \citep[KINGFISH,][]{kenn_11}
also find that the nonthermal spectrum in galaxies with a higher SFR
surface density ($\Sigma\, {\rm SFR}$) is flatter \citep{Taba17}. This shows that the bulk of the CREs population is younger and more energetic as a result of massive star formation activities. Combined with
polarization observations, the synchrotron emission reveals that the
 magnetic field becomes stronger with SFR too \citep{niklas, chyzy,taba_13,Heesen}. As suggested by \cite{Taba17}, the stronger magnetic field together with a flatter synchrotron spectral index is due to winds/outflows driven by turbulent magnetic field/cosmic rays, which is called {\it nonthermal feedback}. This was confirmed by the VLA Cloud-Scale Survey of Star Formation and Feedback in M33, revealing that it is rooted in star-forming regions with a turbulent/tangled magnetic field: Here, injected CREs are scattered off from the many pitch angles of the tangled fields. These high-energy particles then insert the pressure gradient between the disk and halo, resulting in winds and outflows \citep[][see Fig.~\ref{M33}]{taba_22}.  Advection of CREs inferred through radio scale-height studies of edge-on galaxies \citep{krause,heesen_16,Misk,Xu} and detailed in theoretical works \citep[e.g.,][]{pfrommer} may have the same nonthermal origin which should be explored. 
 As part of the MeerKAT International GHz Tiered Extragalactic Exploration, the RC observations of highly star-forming galaxies at $1.5<z<3.5$ show that the nonthermal feedback has been even more efficient at cosmic noon \citep{taba_25}. Have galaxies, during their evolution, been even more influenced by this kind of nonthermal feedback than the thermal winds proposed by models \citep[e.g.,][]{Schaye}? This is an open question that requires studies of energy/pressure balance in galaxies over cosmic time; a science focus of the SKAO Extragalactic Continuum Working Group \citep{ghasemi,taba_24}.

\subsection{ISM/IGM energy balance, technical challenges, and importance of SKAO}
\begin{figure*} 
\centering
\includegraphics[width=0.47\hsize]{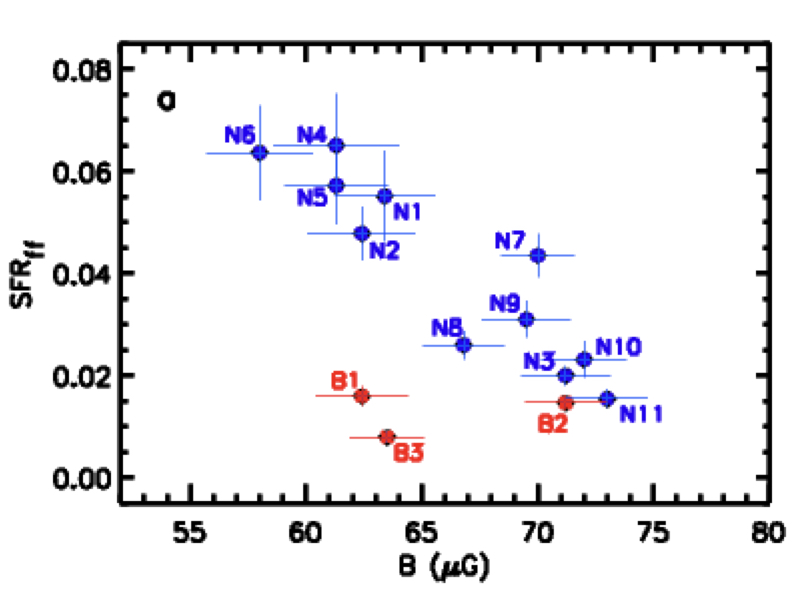}
\includegraphics[width=0.47\hsize]{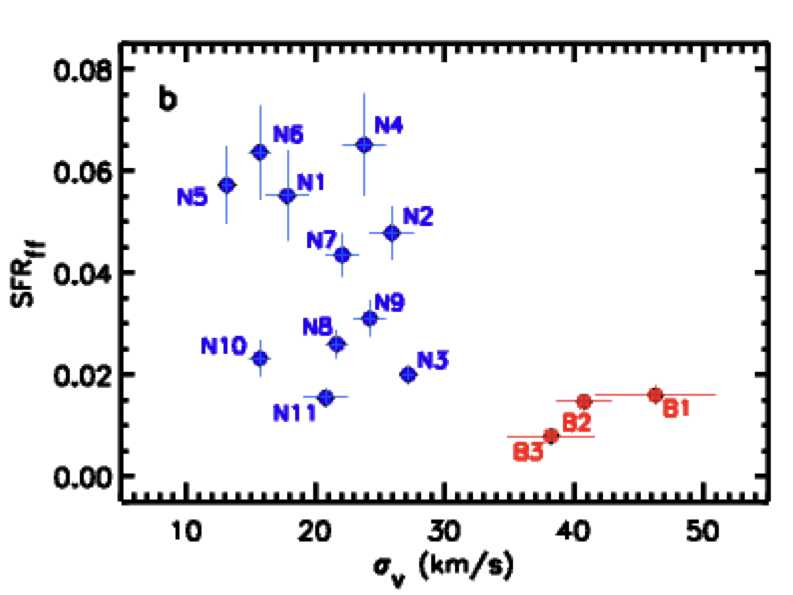}
  \caption{Role of nonthermal processes on star-formation rate per free-fall ($\rm SFR_{ff}\equiv \Sigma\,{\rm SFR}/\Sigma\,{\rm H_2} \times \tau_{\rm ff}$ with $\Sigma\,{\rm SFR}$ and $\Sigma\,{\rm H_2}$ the surface densities of the SFR and molecular gas, respectively, and $\tau_{\rm ff}= 4.7\,(\frac{M_{H_2}}{10^6\,M_{\odot}})^{0.25}$ is the the free-fall timescale of molecular gas) in giant molecular cloud associations in the circumnuclear ring of NGC1097. {\it Left-} ${\rm SFR_{ff}}$ vs equipartition magnetic field strength. {\it Right-} ${\rm SFR_{ff}}$ vs turbulence traced by CO(2-1) velocity dispersion. Points are clouds with CO(2-1) narrow (blue) and broad (red) line widths \citep[see][]{taba_18}.}
\label{N1097}
\end{figure*}

Observations show that the nonthermal pressure inserted by cosmic rays and magnetic fields can be higher than the thermal gas pressure in the ISM of spirals \citep{beck07,Taba_08,Basu}, dwarf irregulars such as the Magellanic Clouds \citep{Hassani}, and the Galaxy Center \citep{Mazoochi}. The nonthermal processes which involve magnetic fields and cosmic rays can control accretion from
the IGM \citep[e.g.,][]{owen,Rint}, collapse/fragmentation in the ISM \citep[e.g.,][]{pillai}, and the star formation rate efficiency of molecular cloud associations \citep[][see Fig.~\ref{N1097}]{taba_18}. The effect of nonthermal processes on the cosmic evolution of galaxies is a pressing question. These processes are best traced through
the synchrotron continuum radiation observed at radio frequencies. However, the
RC emission from galaxies is not purely due to synchrotron radiation. Particularly in star-forming regions, it can be due to the free-free
radiation of thermal electrons by about 50\%  or even higher at 6 GHz \citep[up to 30\% at 1GHz,][]{taba_13,taba_22}. Hence, separation of the thermal and nonthermal emission is the first step in these studies. In global galaxy studies, modeling the spectral energy distribution of the RC emission has been an ideal separation method \citep{condon_92,Taba17}. In resolved studies, this method requires a consistent sensitivity at different frequencies, which is a technical challenge. Hence, results based on this method have been mainly focused on only bright star-forming regions \citep[e.g.,][]{Murphy_10}, have assumed fixed or weighted spectral index \citep[e.g.,][]{Dignan}, or have been limited to a coarse resolution \citep[][]{West}.   
As a consequence, most of our knowledge of the distribution of the pure synchrotron emission, CREs energy index, and magnetic field
strength in the ISM of galaxies is based on a complicated technique which uses a physically motivated non-radio tracer for the free-free emission such as recombination lines like de-reddened H$\alpha$ emission \citep[][]{taba_07,taba_13b,taba_18,taba_22,Hassani}. 
Sensitive observations with SKAO in synergy with optical/NIR observations should make it possible to overcome this historically technical challenge in studies of high-z galaxies.

\section{Resolving the ISM/IGM components in the distant universe with SKAO}

Thanks to its unprecedented sensitivity, the SKAO  will transform our knowledge in the field of galaxy formation and evolution. In this section, we investigate the SKA Bands\,1 and 2 capability to map the RC and  HI 21-cm emission from main-sequence galaxies at high-z. This is needed to address the most pressing question: 
\begin{itemize}
    \item What is the role of the thermal and nonthermal pressures in the formation of the ISM structures over cosmic time? 
    \item Do the thermal fraction and the synchrotron spectral index of star-forming galaxies change with the redshift?
    \item How does the gas content of the ISM evolve with the redshift? 
\end{itemize}
  Addressing these questions is important for energy balance and galaxy evolution studies. In the following, we outline simulations performed to showcase the capability of the SKAO in mapping the energetic ISM components traced by the thermal and nonthermal RC emission (Sect. 4.1) and the HI gas surface brightness at an angular resolution of 0.6" in distant ($z>0.1$) galaxies.

\begin{figure*} 
\centering
\includegraphics[width=0.99\hsize]{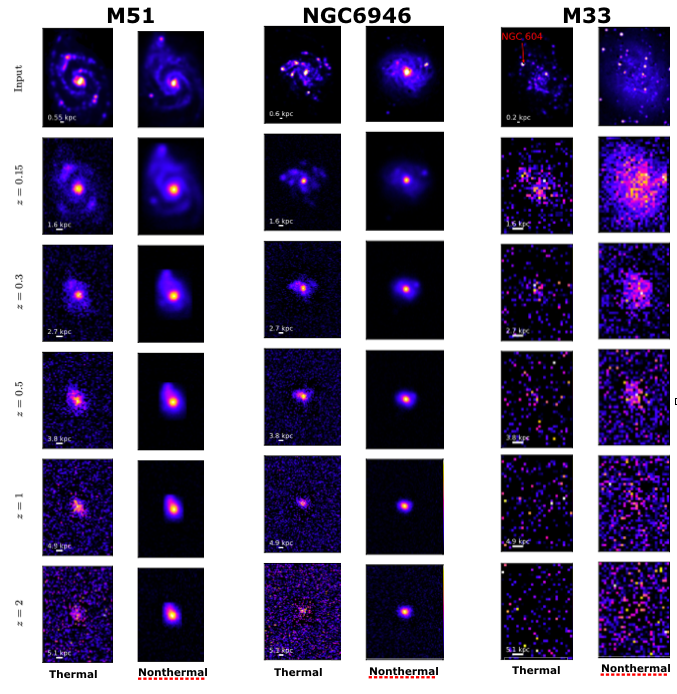}
      \caption{Simulated surface brightness maps of thermal and nonthermal emission from analogs of M51, NGC6946, and M33 galaxies at different redshifts z=0.15, 0.3,0.5, 1, and 2 simulated for the SKA AA4 at 0.6" resolution at the observed frequency of 1.4 GHz \citep{ghasemi}.}
\label{maps}
\end{figure*}

\begin{figure*} 
\centering
\includegraphics[width=0.45\hsize]{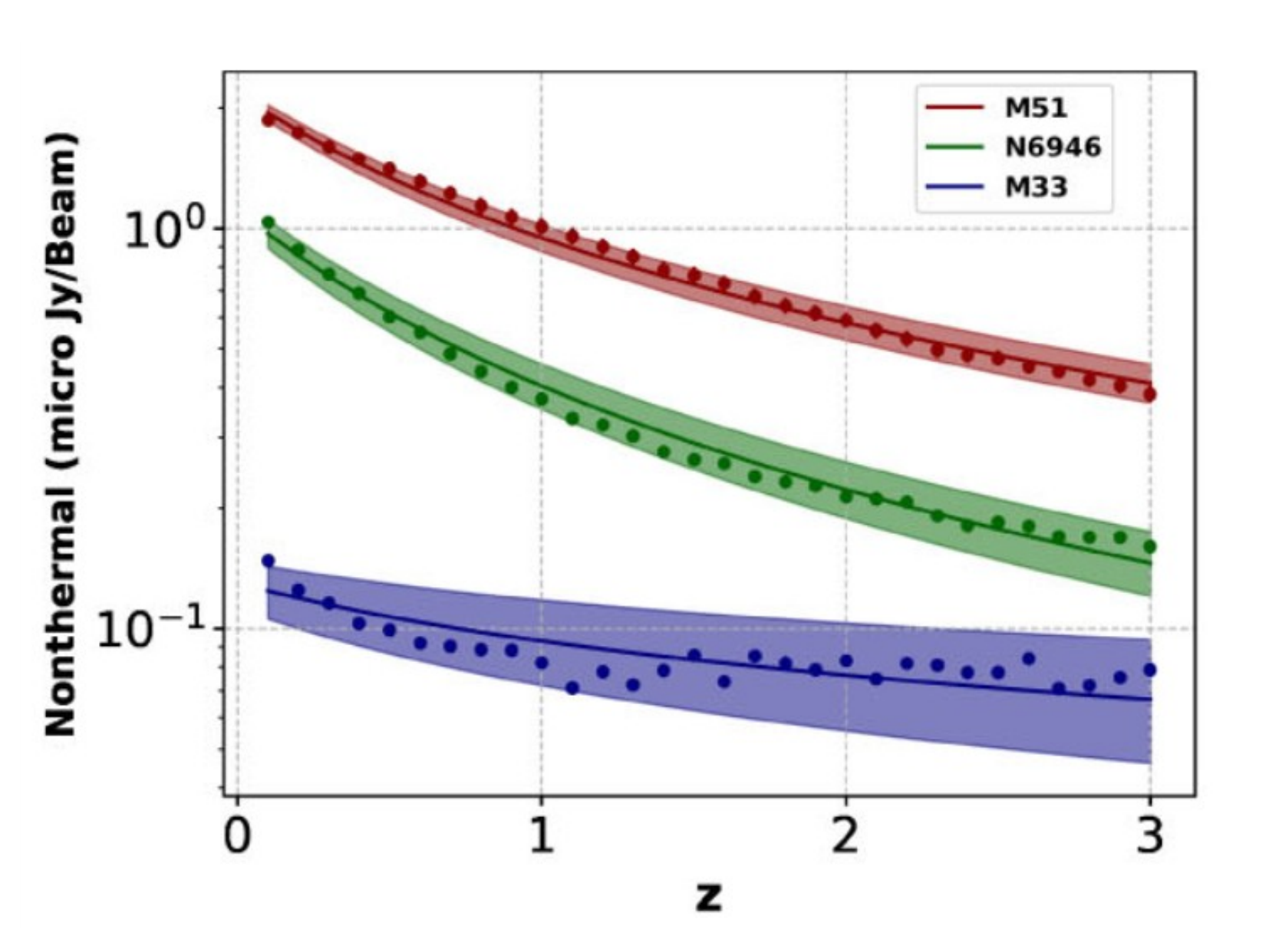}
\includegraphics[width=0.45\hsize]{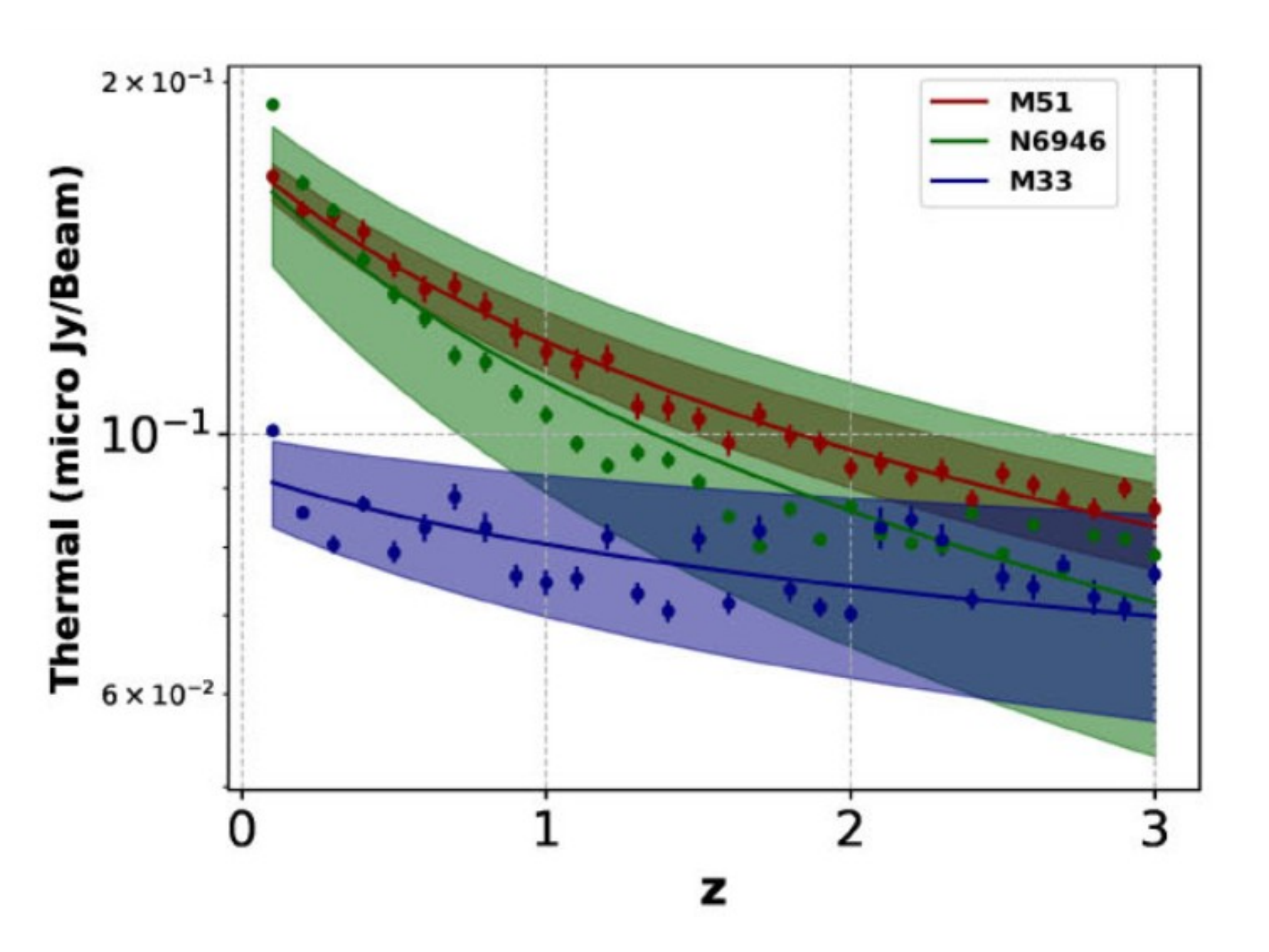}
\includegraphics[width=0.45\hsize]{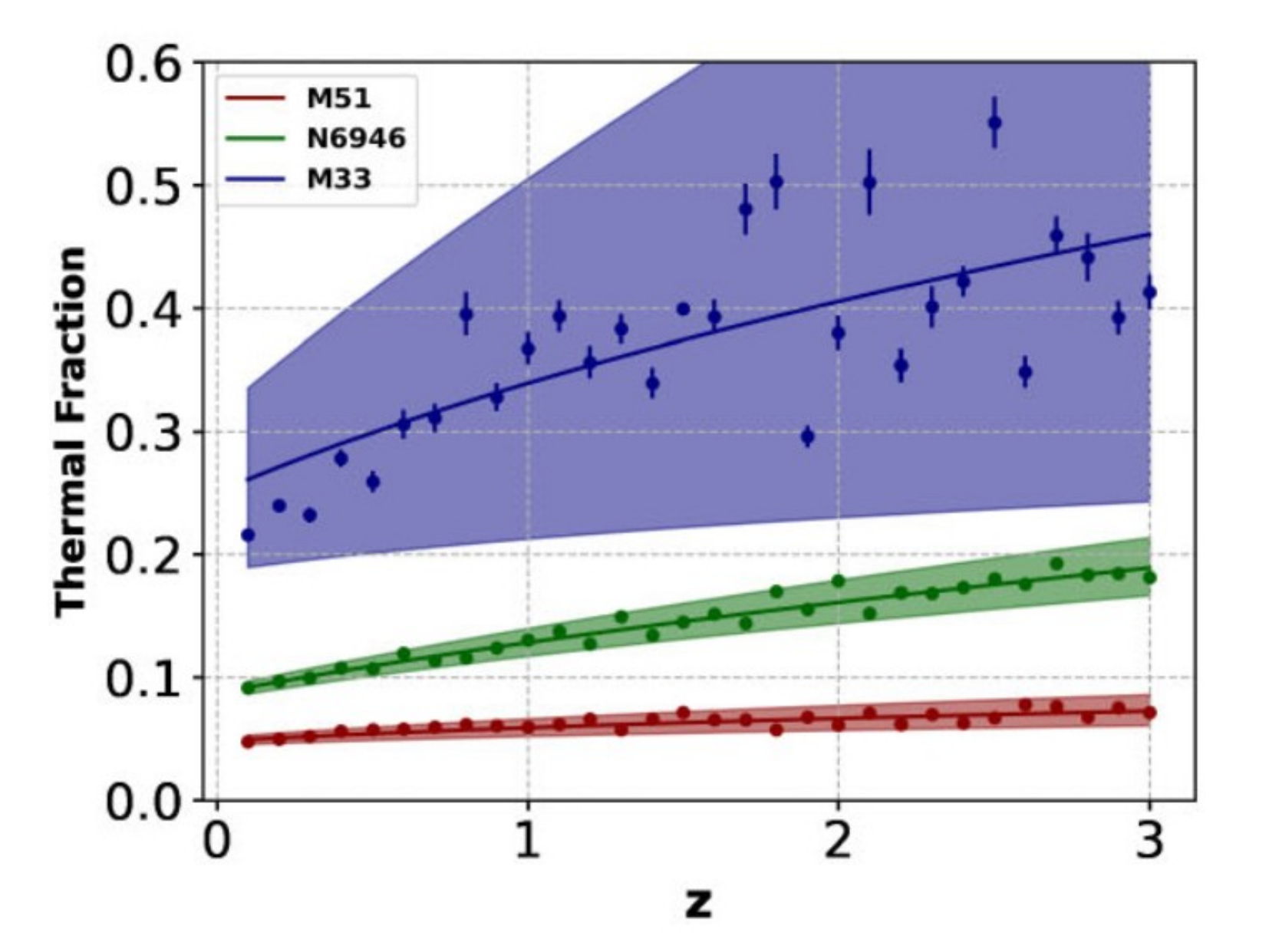}
\includegraphics[width=0.45\hsize]{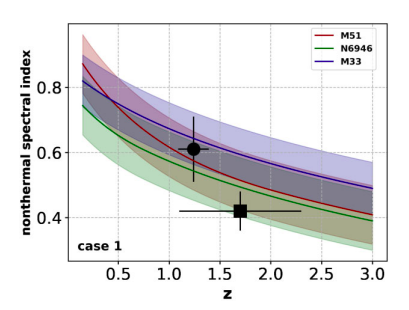}
      \caption{{\it Top:} Redshift evolution of the nonthermal (left) and thermal (right) mean surface brightness at the observed frequency of $\nu_2$ = 1.4 GHz (equivalent to a rest-frame frequency of 1.6–5.6 GHz at $0.15 \leq z \leq 3$) and the thermal fraction (bottom-left).  Shaded regions show the 3$\sigma$ uncertainties of the fits of the form $a(1 + z)b$  \citep[see][for more details]{ghasemi}. Also shown is the evolution of  $\alpha_{\rm nt}$ with redshift ({\it bottom-right}). Shadowed bands show errors in the spectral indices. Points indicate published measurements by \cite{Murphy_17} (circle) and \cite{tisan} (square).}
\label{AlphantHistfth}
\end{figure*}

\subsection{Tracing back the RC emission from the main sequence galaxies at high redshifts with AA4 and AA$^{\star}$}

\begin{figure*} 
\centering
\includegraphics[width=0.7\hsize]{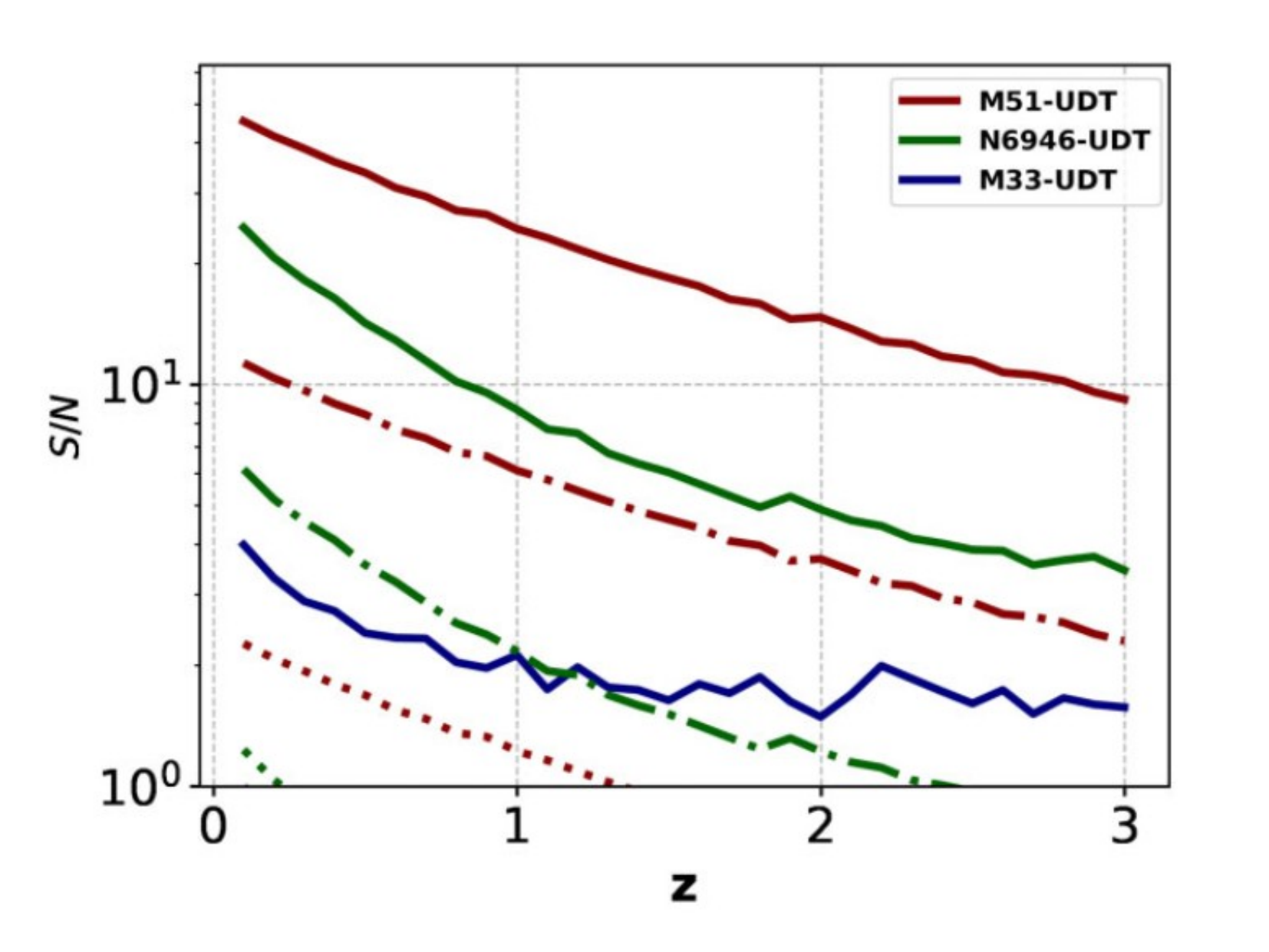}
      \caption{Mean S/N based on the SKA AA4 band 2 UDT with the rms noise level of $\sigma=0.05\,\mu$Jy (solid), DT with $\sigma=0.2\,\mu$Jy
(dashed-dotted), and WT with $\sigma=1\,\mu$Jy (dotted) surveys at the observed frequency of 1.4 GHz (equivalent to a rest-frame frequency
of 1.6–5.6 GHz at $0.15 \leq z \leq 3$). }
\label{snr}
\end{figure*}

 In \cite{ghasemi}, we obtained an expression for the cosmic evolution of the thermal and
nonthermal RC surface brightness from star-forming galaxies, taking into account
account the correlation between the SFR and the thermal/nonthermal
RC emission \citep{Murphy_11} and the evolution of the SFR \citep{schrei}:
\begin{equation}
    I(z) = I(0) \frac{{\rm SFR}(z)}{\rm SFR(0)}\frac{D^2(0)}{D^2_L(z)} (1+z)^{1-\alpha},
\end{equation}
where $I(z)$ is the observed surface brightness of the galaxy at redshift $z$. In the above relation, $I(0)$, SFR(0), and $D(0)$ are the surface brightness, SFR, and distance of the galaxy in the local universe ($z\simeq0$), whereas SFR$(z)$ and $D_L(z)$ are the SFR and the luminosity distance of the source at redshift $z$. For thermal emission $\alpha=\alpha_{\rm th}$ ($=0.1$ in optically thin condition) and for nonthermal emission\footnote{The definition of $I_{\nu}\propto \nu^{-\alpha}$ is used in these expressions.} $\alpha=\alpha_{\rm nt}$ which becomes flatter with increasing the SFR surface density \citep{Taba17} or specific SFR \citep{taba_25} in galaxies. 
%
%
%
Using Eq\,(1), we simulated back the RC maps of M51, NGC6946, and M33 analogs at high-z at an observed frequency of 1.3\,GHz. These galaxies were chosen because 1) they refer to a wide range of SFR in the main-sequence population of galaxies and 2) their maps of the thermal and nonthermal RC are available \citep{taba_13,taba_13b}.  We investigated the detectability of the ISM structures in the three-tiered surveys of wide tier (WT), deep tier (UT), and Ultra-deep tier (UDT) proposed by \cite{prand}\footnote{For an update, see \cite{Prandoni01.2026.SKA}.} at an angular resolution of $\theta=$0.6" \citep{ghasemi}. Here, we revisit those simulations taking into account the SKA-AA4 capabilities at $\theta=$0.6" and 0.9". Maps were produced at redshift intervals of 0.1 for $0.15<z<3$, taking into account the angular diameter evolution. Figure~\ref{maps} shows the resulting maps at $z=0.15, 0.3, 0.5, 1, 2$ as observed by AA4-Band2 at the UDT rms sensitivity of 0.05$\mu$Jy/beam. Our simulations show that M51 analogs can be mapped in both UDT and DT surveys up to cosmic noon ($z=2-3$), while M33 analogs can hardly be detected at $z>0.3$ with AA4 (Fig.~\ref{snr}).

\begin{figure*} 
\centering
\includegraphics[width=0.8\hsize]{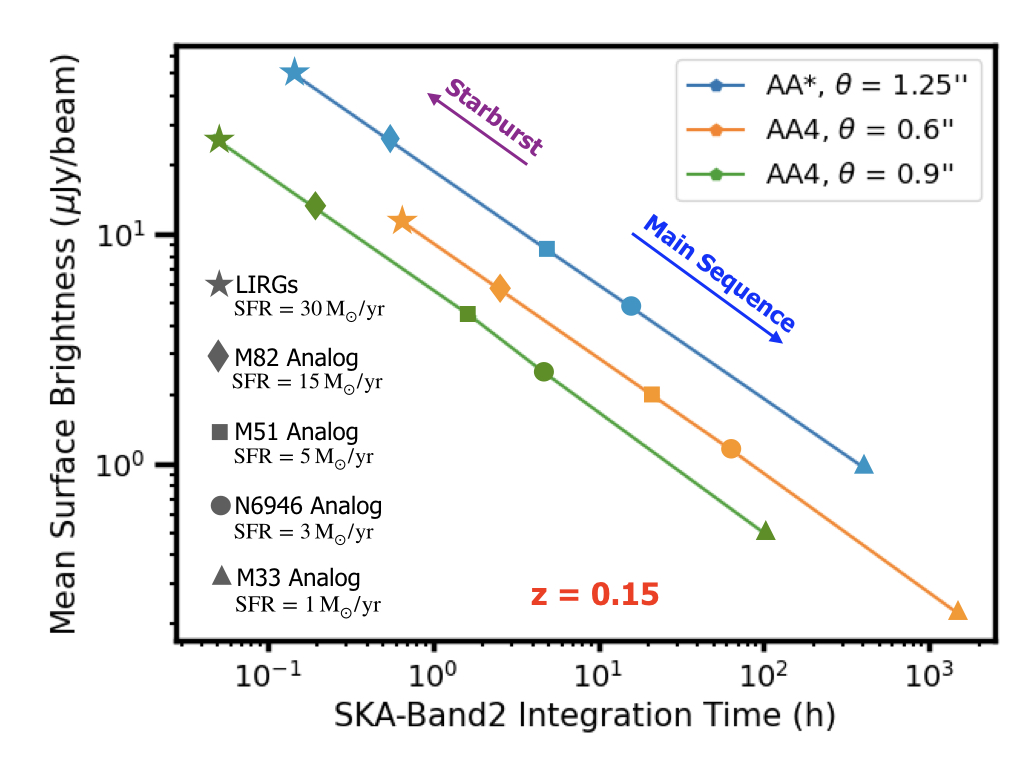}
\includegraphics[width=0.8\hsize]{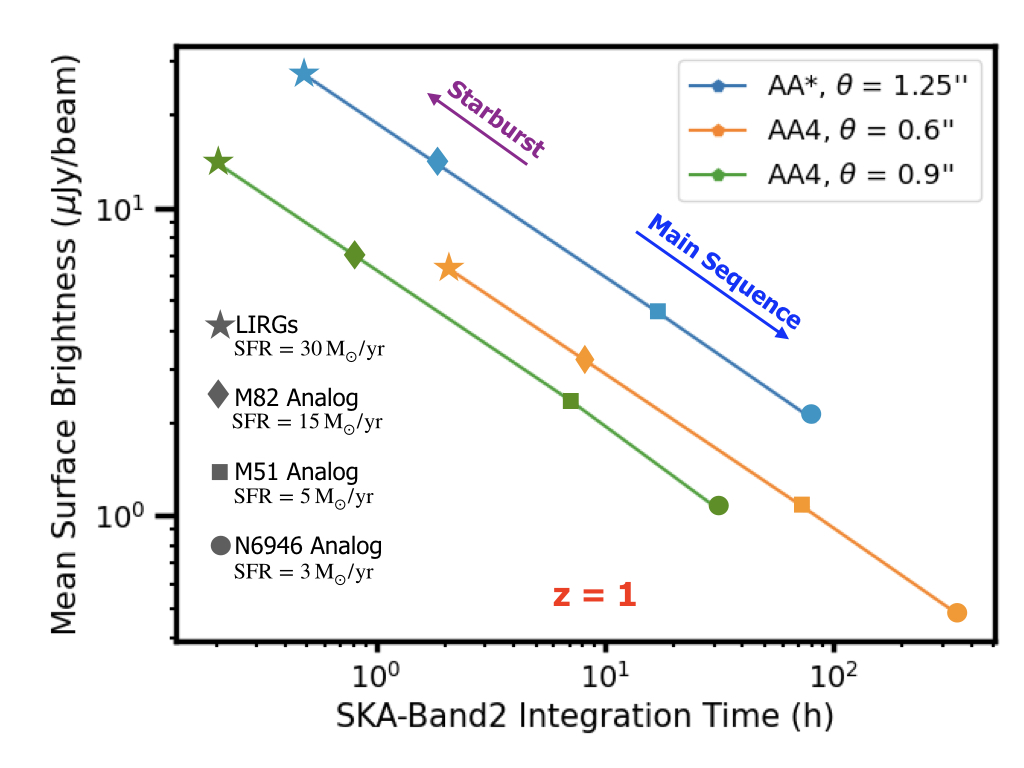}
      \caption{{\it Top-}  Mean radio continuum surface brightness of star-forming galaxies at $z=0.15$ ($5\,\sigma$ detection) vs the SKA Band2 integration time observed with AA$^{\star}$ at $\theta=1.25"$ and AA4  at $\theta=0.6"$ and $0.9"$ angular resolutions. {\it Bottom-} Same as in the top for galaxies at $z=1$ excluding the case for a low surface brightness galaxy such as M33 (M33 Analog). Solid lines just indicate the observational setups.}
\label{AA}
\end{figure*}

Figure \ref{AA} shows the time needed to detect the mean surface brightness of the RC emission in these galaxies (5$\sigma$ detection). In this plot, M82 analogs and LIRGS are also shown, assuming that a linear RC-SFR correlation holds for these galaxies. The angular scale of 0.6" allows the study of the ISM structures at scales down to $<1$\,kpc (5\,kpc) at $z=0.15$ ($z=2$) which include star-forming associations, spiral arms, and clumpy structures in disks. At 0.9" resolution, the ISM structures can be resolved down to 2 to 8\,kpc at $0.15<z<2$, including spiral arms and inner and outer galaxy disks.  We note that the observation time decreases by about one order of magnitude by degrading the angular resolution from 0.6" (Briggs=-1, tapering=0.267") to 0.9" (Briggs=0, tapering=0.267").

Already in its deployment phase (AA$^{\star}$), the SKA observatory can map high-z star-forming galaxies including low-surface brightness galaxies such as M33 at feasible observing times but at a poorer angular resolution of 1".25 according to the SKA Science Calculator (setting Briggs=0 and tapering=0.267"). As shown in Fig.~\ref{AA}, these observations need a longer integration time by about a factor of 3 to 4 compared to the AA4 observations at the slightly better angular resolution of 0.9". Hence, studying the ISM/IGM structures,  AA$^{\star}$ can act more effectively at lower redshifts ($z\lesssim  0.5$), where its angular resolution ($\theta=$\,1".25) allows mapping  at least 4 resolution  elements along the major axis of a galaxy.   



\subsubsection{Mapping  synchrotron polarization in high-z galaxies with SKAO}

\begin{figure*} 
\centering
\includegraphics[width=0.95\hsize]{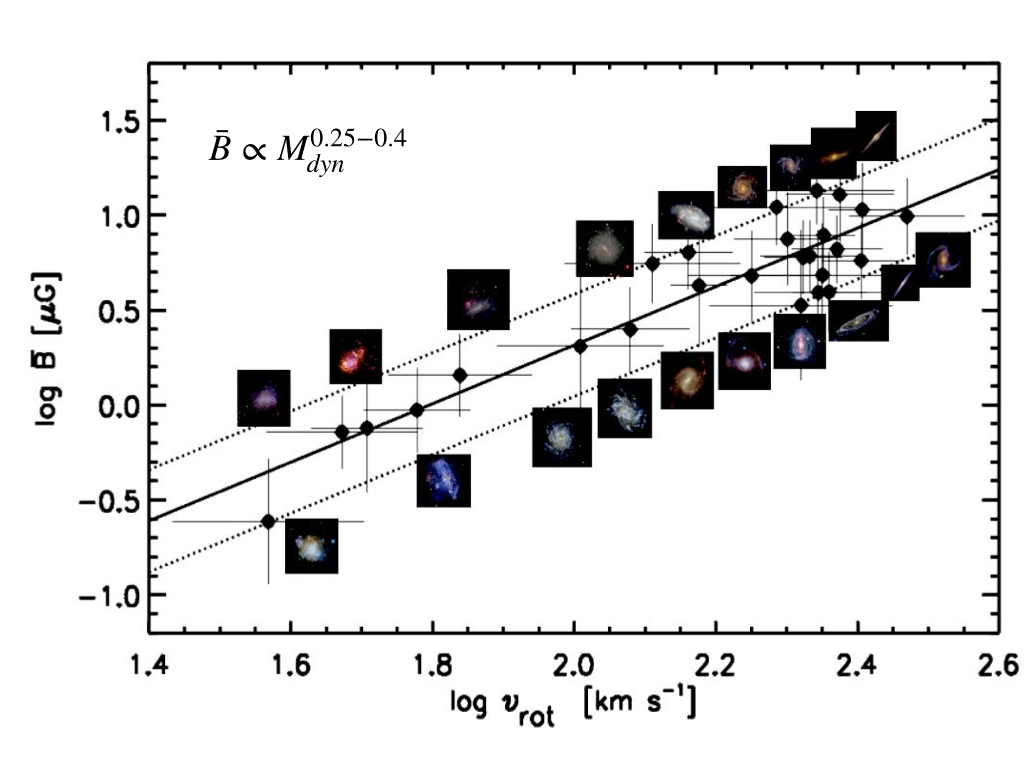}
      \caption{Faster rotating and/or more massive isolated galaxies have stronger large-scale magnetic fields ($\bar{B}$) traced by synchrotron polarization \citep{taba_16}.  The scatter at the higher mass end (rotation velocities $v_{\rm rot}>\,210$\,km\,s$^{-1}$) is linked to an excess SF activity (enhancing $\bar{B}$) and ejection from super-massive black holes (reducing $\bar{B}$) in TNG50 simulations \citep{rad}.  This corelation shows that the large-scale ordered field in nearby galaxies is dominated by anisotropic turbulent field. Full polarization surveys with SKA will extend this study to larger galaxy samples in both local and high-z universe.}
\label{Bmass}
\end{figure*}

Whereas the nonthermal synchrotron emission traces  the total magnetic field strength, its polarization traces the ordered magnetic field\footnote{The observed magnetic field in galaxies also has a random or turbulent component which often prevails over the ordered one \citep[e.g.,][]{beck_15}.} projected on the sky plane.  
Full polarization RC observations unveil the distribution and structure of the ordered field in galaxies. However, our current knowledge is mainly limited to only small samples of nearby galaxies at kpc scales. The SKA reference surveys  \citep{prand} and those discussed above will also map the synchrotron polarization, provided that they are carried out in full polarization. Based on the experience with SKA pathfinders, the rms noise in the Stokes-Q and -U maps is naturally lower than that in the Stokes-I maps. This ensures detection of polarization signals at sensitivity levels discussed above, at least  at rest-frame frequencies of $\nu\geq$\,3\,GHz (for a conservative assumption of possibly intense depolarization at lower rest-frame frequencies in high-z galaxies). This is ideally possible at rest-frame frequencies (redshifts) which are not much affected by  Faraday depolarization (e.g., rest-frame 10\,GHz which can be covered by Band\,5 up to $z=1.1$ with AA$^{\star}$ and AA4). Such observations are in particular important for expanding our current knowledge of correlations between the large-scale magnetic field and other galaxy properties \citep{van}, including rotation velocity or dynamical mass \citep[][see also Fig.~\ref{Bmass}]{taba_16}, to higher redshifts. We note that multi-band observations are needed to infer the 3D structure of the ordered field through Faraday rotation measure analysis. Multi-band observations will also help to separate the thermal/nonthermal RC components through modeling the radio spectral energy distribution  \cite[see chapters by][for other important applications of the SKA multi-frequency surveys]{FangxiaAn01.2026.SKA,Prandoni01.2026.SKA}.

\subsection{Tracing back the HI 21-cm surface brightness of M51 analogs at high redshifts with SKAO}\label{HIsimu}
\begin{figure*} 
\centering
\includegraphics[width=0.7\hsize]{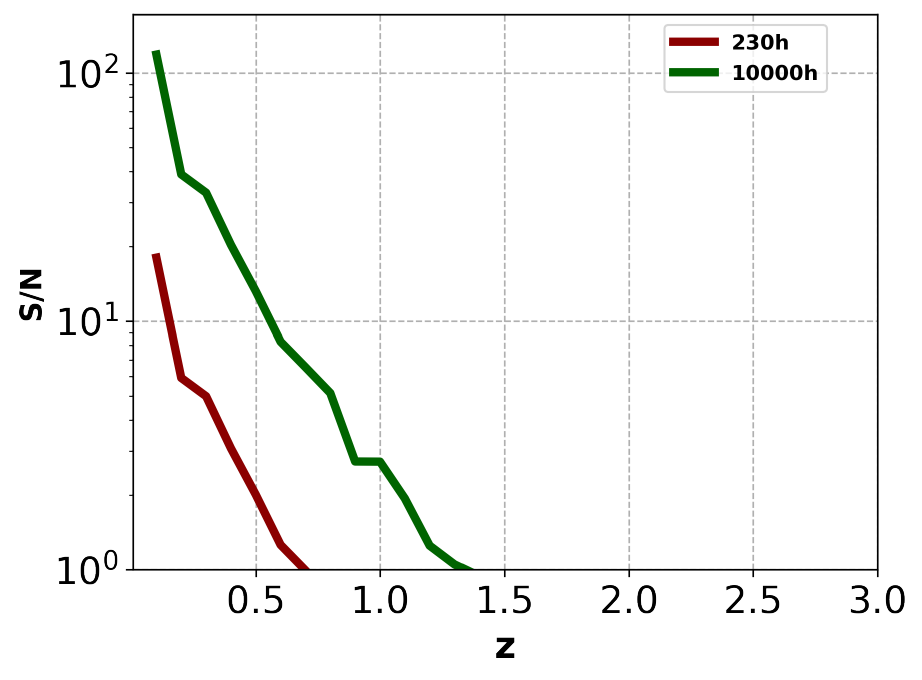}
      \caption{The mean S/N of simulated HI moment 0 maps of M51 analogs vs redshift.} 
\label{means}
\end{figure*}

To assess the SKA capability to map the cool (neutral) gas content of galaxies over cosmic time, we also simulated the HI 21-cm line emission in an M51 analog at high-z. 
Using a realistic SKA sub-array configuration in the CASA task {\it simobserve}, the AA4 visibilities were simulated for Bands\,1 \&\,2 and a channel width of $13$\,kHz. The HI cubes were generated for each redshift interval of 0.1 over $0.1\leq z \leq 3$. An rms noise level of 5\,$\mu$Jy per channel was injected as sky level.   
%
Figure~\ref{means} shows the resulting  mean S/N of the HI maps at different redshifts. The mean atomic ISM can be mapped back to $z\simeq0.4$ at 2.5\,$\sigma$ level already after 230\,h of observation in the M51 analogs. This is a significant improvement to  observations possible with available telescopes, which is limited to $z\lesssim\,0.1$. However, we need a much longer integration time of $\simeq$10,000\,h to map HI up to $z\simeq1$ with the same signal to noise ratio of 2.5. 
This agrees with the estimates presented by \cite{Lelli01.2026.SKA} who proposed a very deep ($\simeq$10,000\,h) observation in a cosmological field with SKA-AA4.  We note that resolved HI 21-cm
observations, which are also important to study the rotation curves and dark matter content of galaxies beyond $z\simeq$\,1, require a higher sensitivity than that provided by AA4.

\section{Synergies with ALMA, VLT/JWST, Euclid}

    Studying the ISM/IGM structure formation and energy balance requires a complete view of different phases of the ISM, including the molecular gas phase. ALMA surveys can be used or performed in synergy with SKAO observations. Furthermore, as discussed earlier, to separate the thermal and nonthermal emission, rest-frame maps of optical/NIR recombination line data are needed, which can be provided in synergy with VLT/JWST observations. Moreover, Euclid can help to address the evolution of the IMF and stellar population over cosmic time. A more detailed specification of possible synergies with optical observations is provided by \cite{Prandoni01.2026.SKA}.

\section{Summary}

As members of the ISM/IGM focus group of the SKAO Extragalactic SWG, we have studied the capability of the SKA-AA4 (and AA$^{\star}$) to map the RC and HI surface brightness in distant star-forming galaxies.  Our results are summarized as follows:

\begin{itemize}
    \item AA4 can map the RC emission from M51 analogs at z=1 at 0.9" angular resolution in less than 10 hours in Band\,2, reaching a 5$\sigma$ detection. Deeper Band\,2 observations ($<100$\,h) will resolve structures in M51 and NGC6946 analogs at cosmic noon ($\theta\leq$0.6"). Galaxies with higher SFR can be mapped in a shorter time and at higher-z.   

    \item Low-mass galaxies like M33 can be mapped at $z<0.3$ at 0.6" at 3$\sigma$ level. At a degraded resolution of 0.9", they can be mapped in about 100\,h at $z=0.15$.

    \item HI simulations indicate that SKA-AA4 can map the HI gas in M51-like galaxies beyond the local universe (up to $z=0.4$) in about 230\,h. Much longer observations ($\simeq$ 10,000\,h) are needed to resolve atomic gas in the ISM/IGM of galaxies up to $z\simeq1$. 

    \item Our RC simulations show that the thermal fraction increases slightly with redshift while the synchrotron spectrum flattens, confirming that cosmic ray electrons are more energetic at higher redshifts due to the evolution of SFR, which can have consequences in energy balance and structure formation in early galaxies. 

    \item The proposed SKA RC  surveys and observations will also shed light on the large-scale magnetic fields  in high-z galaxies,  provided that they are carried out in full polarization. This will further allow us to understand the mechanisms that set the degree of order of the magnetic field structure in galaxies when combined with gas kinematics information  with SKA-AA4 HI surveys at least up to $z=1$.

\end{itemize}

{
\section*{Acknowledgement}
FST acknowledges the support from the DYNAVERSE Cluster of Excelle (Cologne – Bonn) which is
Funded by the Deutsche Forschungsgemeinschaft (DFG, German Research
Foundation) under Germany's Excellence Strategy – EXC 3037 – 533607693. MGN acknowledges the support from the CAS Talent program and the Xinjiang Tianchi Talent program. JM acknowledges financial support from the Spanish grant PID2023-147883NB-C21, funded by MCIU/AEI/ 10.13039/501100011033, as well as support through ERDF/EU, and from the Severo Ochoa grant CEX2021-001131-S funded by MCIN/AEI/ 10.13039/501100011033.
}
\bibliographystyle{abbrvnat-maxbibnames4}
\bibliography{chapter} 

\end{document}